# Optimizing Coherence Suppression in a Laser Broadened by Phase Modulation with Noise

Jonathan M. Wheeler, Jacob N. Chamoun, and Michel J. F. Digonnet

*Abstract*— **Phase noise can be modulated onto the output of a laser with an electro-optic phase modulator (EOM) to create a highly incoherent broadened source with low intensity noise. This technique leaves a small but finite fraction of the coherent carrier power that can be highly detrimental in applications requiring incoherent light. This paper shows that the carrier suppression in a laser broadened by this technique can be calculated for an arbitrary noise probability density function. The carrier suppression can be varied experimentally by adjusting the noise voltage standard deviation $Gv_\sigma$ and saturation voltage $V_{sat}$ of the amplifier that amplifies the noise source that drives the EOM. Simulations show that suppressions better than -30 dB are attainable for reasonable tolerances in $Gv_\sigma$, for example an EOM with a $V_\pi$ of 4.7 V, a $V_{sat}$ of 6.3 V, and $Gv_\sigma/V_\pi = 0.73 \pm 0.03$. Even greater tolerances can be achieved with a higher $V_{sat}$ (21 V), lower $V_\pi$ (3.5 V), and $Gv_\sigma/V_\pi \approx 3.2 \pm 2.3$. This model aids in selecting these parameters such that the carrier suppression is resilient to fluctuations in these voltages due to temperature variations and aging of the components. The model predictions are validated by testing a fiber optic gyroscope interrogated with a broadened laser for various $Gv_\sigma$ and $V_{sat}$ and inferring the carrier suppression from its measured noise. A -44-dB carrier suppression was observed for a nonlinear amplifier with $Gv_\sigma = 3.43$ V, $V_{sat} = 6.3$ V, and an EOM with $V_\pi = 4.7$ V, in agreement with predictions.**

*Index Terms*—**Electrooptic modulators, optical interferometry, optical coherence tomography, phase modulation, Sagnac interferometers, fiber optic gyroscope**

## Nomenclature

| | |
|---|---|
| $f_c$ | fraction of total power in residual carrier |
| $v_n(t)$ | noise signal before final amplifier |
| $V_N(t)$ | noise signal incident on EOM input |
| $p(v_n)$ | probability density function of $v_n(t)$ |
| $h$ | transfer function of final amplifier |
| $P(V_N)$ | probability density function of $V_N(t)$ |
| $V_\pi$ | required voltage to induce a half-wave shift at EOM |
| $v_\sigma$ | standard deviation of $p(v_n)$ |
| $v_\mu$ | mean of $v_n(t)$ |
| $V_{sat}$ | maximum output voltage of final amplifier |
| $x_{a,b}$ | lower and upper limits of amplifier linear regime |
| $\alpha_i$ | attenuation of the $i$th attenuator |
| $\Phi_\sigma$ | standard deviation of the phase shift, $\Phi_\sigma = \pi G v_\sigma / V_\pi$ |
| $N$ | the number of modeled discrete frequencies |
| $\Phi_p$ | peak phase shift of a discrete frequency, $= (2/N)^{1/2}\,\Phi_\sigma$ |

## I. Introduction

In certain classes of two-wave interferometers, in particular for laser imaging [1], rotation sensing [2], and optical coherence tomography [3], spurious signals arising from imperfections such as scatterers interfere coherently with the signal of interest and add noise and/or drift to it. These applications benefit from the use of incoherent light, which eliminates these interferometric effects outside of a small spatial region. For example, some LIDAR use incoherent light to “see through” atmospheric turbulence and scattering from aerosols to measure displacements of distant objects [4]. In medical imaging instruments, incoherent light is used to image μm$^3$ volumes inside a tissue that would otherwise be difficult to resolve with coherent light due to background reflections from surrounding tissue [3]. Fiber-optic gyroscopes (FOGs) measure extremely small path-length differences (10 fm or less) to infer rotation rates as small as a full turn in 2200 years [2]. This superb precision is achieved by probing the FOG with incoherent light, which significantly reduces the noise and drift arising from spurious interference from light backscattered [5], [6] or polarization-coupled [7] along the FOG’s fiber coil. Stimulated Brillouin scattering, which can limit power scaling of fiber lasers, can also be mitigated by externally modulating the phase of the laser with noise to reduce its coherence [8], [9].

Because the degree of incoherence is proportional to the optical bandwidth, bandwidths of GHz [4], [6], [8], [9] or THz [1]-[3] are often required. Most applications meet this goal by using a pulsed laser or a continuous-wave superfluorescent fiber source (SFS). While SFSs perform well for many applications, their relative intensity noise (RIN) is high, and they have a poor mean-wavelength stability, which in a FOG introduces an error in the measured rotation rate.

Recently, the bandwidth of a laser has been broadened by sending its output through an electro-optic phase modulator (EOM) driven by broadband RF noise [8], [10], [11]. For weak modulation, some of the carrier power is redistributed into symmetric sidebands whose frequency spectra mirror the RF frequency spectrum. As the modulation amplitude is increased, the power in the sidebands is further redistributed into higher-

Manuscript received October 16, 2020. This work was supported by Northrop Grumman Corporation.
J. M. Wheeler is with the Electrical Engineering Department, Stanford University, Stanford, CA 94305 USA (e-mail: jamwheel@stanford.edu).
J. N. Chamoun was with the Applied Physics Department, Stanford University, CA 94305 USA. He is now with PARC, Palo Alto, CA 94304 USA (e-mail: jchamoun@parc.com).
M. J. F. Digonnet is with the Applied Physics Department, Stanford University, Stanford, CA 94305 USA (e-mail: silurian@stanford.edu).

order sidebands, and sidebands convolve with each other: the optical spectrum therefore smoothes out into a Gaussian shape with a broad bandwidth independent of the original laser linewidth. This method has produced bandwidths ranging from ~1.5 GHz [9] to ~10 GHz [8] and 42 GHz [10]. Only a small fraction $f_c$ (~$10^{-4}$) of residual carrier remains [11]. This technique offers the significant benefits of preserving the laser's excellent mean-wavelength stability and producing light with a RIN far lower than the RIN of an SFS.

To be fully effective, the spectrum of such a broadened laser should contain very little of the original carrier. To quantify the negative impact of residual carrier, consider a 25-kHz laser broadened to 25 GHz, with 1% of the power remaining in the carrier. In a FOG, the backscattering noise decreases with increasing optical bandwidth $\Delta\nu$ as $1/\sqrt{\Delta\nu}$ [6,12]. Hence, while the broadband component reduces the noise by $(25\ \text{GHz}/25\ \text{kHz})^{1/2}$, or -30 dB, the noise from the residual 25-kHz carrier is still 1% of the original noise, or ~-20 dB. If the carrier suppression were 0.01%, the noise would be reduced by the full -30 dB. In LIDAR applications, atmospheric and aerosol reflections also introduce spurious two-wave interferometers. When the interferometer's arms differ in length by more than a few coherence lengths $L_c$, the fields do not interfere, and the interferometer output is insensitive to perturbations. For a 1-cm resolution, a bandwidth of a few tens of GHz is needed. A bandwidth of a few THz is required for sub-mm-resolution retina imaging [3]. However, again even just 1% of residual carrier will introduce 1% of coherent reflections and spoil the image quality.

In general, increasing the noise voltage applied to the EOM reduces $f_c$ exponentially [11]-[14]. However, experiments show that at large voltages the carrier suppression degrades [9]-[10], leading to a narrow range of voltages that suppress the carrier optimally. Thus, there is a need to understand why $f_c$ degrades at large voltages, and how to possibly broaden this range. This last feature is important in applications requiring extended operation over a wide temperature range, since the noise source and EOM may operate differently at different temperatures and degrade over time.

This paper derives closed-form equations for the lineshape and residual carrier in a laser of arbitrary linewidth broadened with a phase modulator driven with noise of arbitrary probability density function (PDF). The predicted conditions for maximum carrier suppression are validated by measuring the suppression of both the noise and the drift resulting from the residual carrier in a high-accuracy FOG. The predicted lineshapes are validated by measuring the broadened laser spectrum with an optical spectrum analyzer (OSA). The equation for the lineshape predicts that for strong modulation, the optical bandwidth can be much greater than the RF signal bandwidth of the noise, which is confirmed experimentally. Furthermore, the equation for the carrier suppression predicts three important properties. First, the observed degradation of the carrier suppression at high noise voltages is due to saturation in the amplifiers. Second, this saturation allows for near-perfect carrier suppressions at lower RF power than possible without saturation. Third, the noise voltage tolerance is greatly relaxed if the $V_\pi$ of the EOM is reduced and/or the maximum absolute noise voltage $V_{sat}$ is increased.

## II. Model of Extinction Ratio

In the broadened laser under study (Fig. 1), light from a laser is sent through a high-bandwidth EOM driven with a broadband (several GHz) noise voltage. The noise source consists of a 50-Ω resistor whose Gaussian thermal noise is amplified by a chain of preamplifiers and amplifiers. A high-pass filter is added to remove the $1/f$ noise introduced by the amplifiers below ~100 MHz, which falls partly within the carrier bandwidth. The first seven preamplifiers amplify a relatively small input signal and therefore operate in the linear regime. This first part of the chain therefore generates a white-noise voltage $v_n(t)$ with a Gaussian PDF $p(v_n)$ [10]. In contrast, the final amplifier receives a larger voltage and the voltage signal $V_N(t)$ it delivers to the EOM sometimes saturates at values near $\pm V_{sat}$. As a result, the PDF of $V_N(t)$ is no longer Gaussian, because voltages beyond $\pm V_{sat}$ (in the tails of the PDF) are compressed into lobes at $\pm V_{sat}$. The EOM converts $V_N(t)$ into an optical phase shift $\varphi_N(t) = \pi V_N(t)/V_\pi$, which is applied to the laser. The bandwidth of $\varphi_N$ is limited by the component with the least bandwidth. The EOM outputs a phase-modulated laser with a linewidth of several GHz, plus a small fraction $f_c$ of residual carrier.

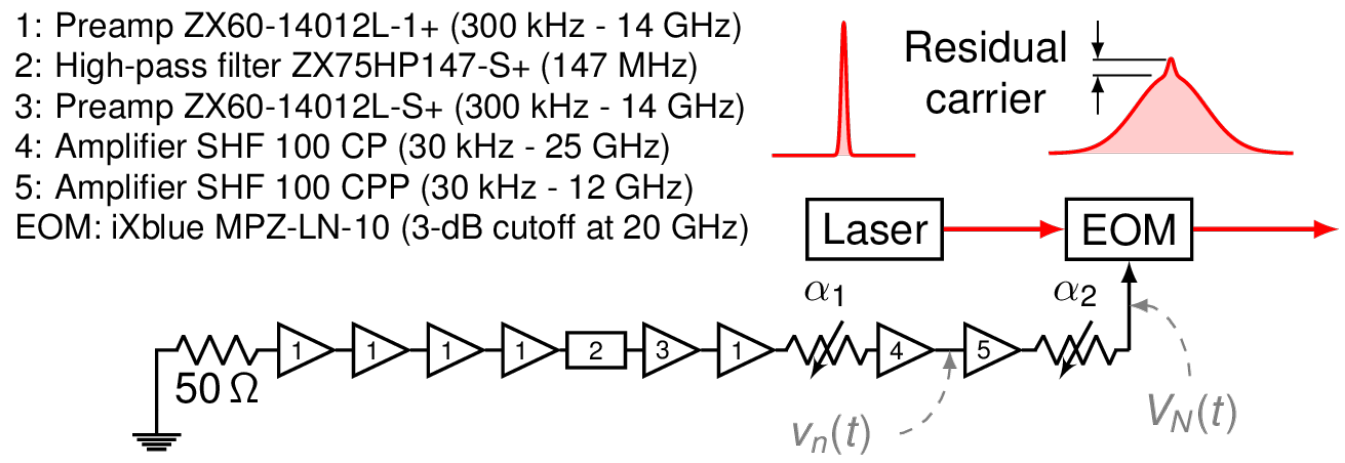

Figure 1. Experimental set up used to broaden the linewidth of a laser to tens of GHz by modulating its phase randomly with an electro-optic modulator driven by broadband noise voltage. This voltage is generated by amplifying the white noise of a 50-Ω resistor with a series of amplifiers.

The standard deviation of the PDF of $v_n(t)$ is defined as $v_\sigma$. Absent any saturation, the standard deviation of the PDF of $V_N(t)$ is then $Gv_\sigma$, where $G$ is the small-signal gain of the last amplifier. A first variable attenuator (attenuation $\alpha_1$) inserted before the last two amplifiers (see Fig. 1) was used to adjust $v_\sigma$, and therefore $Gv_\sigma$. A second attenuator (attenuation $\alpha_2$) was placed after the last amplifier to adjust both $Gv_\sigma$ and $V_{sat}$.

In photonic systems relying on low-coherence light such as LIDAR and FOGs, noise and drift in the measured signal arise from spurious interferometers (e.g., caused by scatterers [4], [7]). To model how the source coherence affects the magnitude of these errors, a spurious interferometer can be modeled as a Mach-Zehnder interferometer (MZI) with an optical-path delay $\tau = n\Delta L/c$, where $\Delta L$ is the length difference between the two arms, $n$ is the index of the arms, and $c$ is the speed of light. $\Delta L$ represents, for example in a LIDAR, twice the distance between the target (which produces the main reflected signal of interest) and a scatterer (which produces a spurious reflection). The MZI beam splitters are lossless 50% couplers. The interferometer is interrogated by a laser broadened to a perfect Gaussian lineshape with a 10-THz linewidth either with no residual carrier (orange curve in Fig. 2(a)) or with 10% of residual carrier power (blue curve).

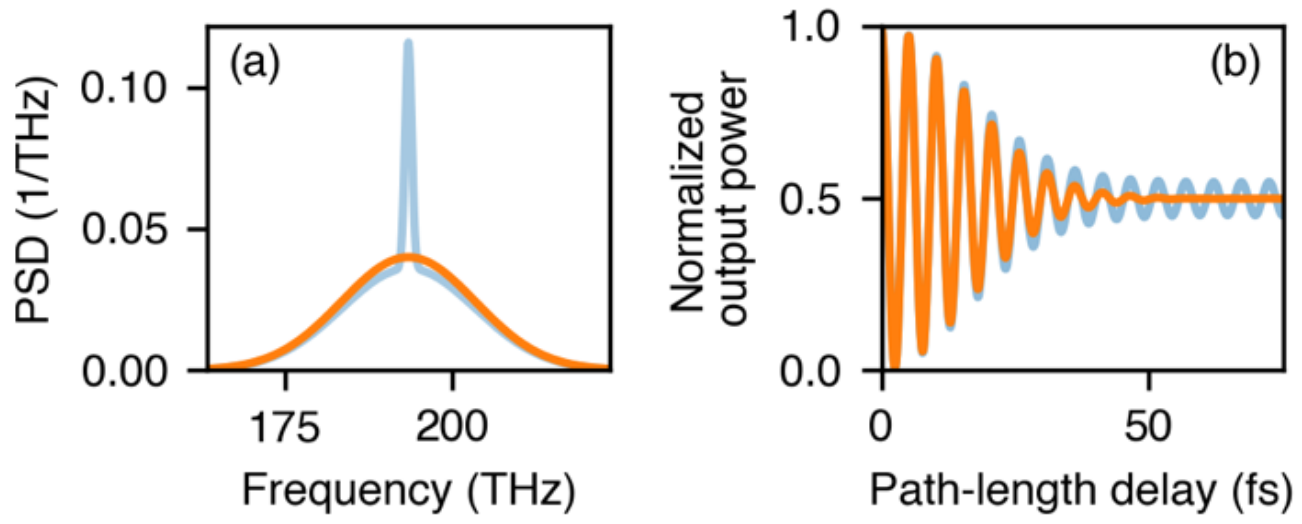

Figure 2. (a): Optical spectrum of a laser homogeneously broadened to a Gaussian lineshape with a 10-THz linewidth with no residual carrier (orange curve) or with 10% residual 500-GHz carrier (blue curve). (b): Average interferometric output power of a two-wave interferometer driven by each of these lasers as the path-length mismatch is increased.

The average output power of the MZI interrogated with each laser is plotted as a function of the optical-path delay $\tau$ in Fig. 2(b). As $\tau$ is increased, the output exhibits interference fringes with an envelop $\Gamma(\tau)$ (or contrast) of diminishing amplitude. For the laser with perfect carrier suppression (orange curve in Fig. 2(a)), this envelop is given by [19]

$$\Gamma(\tau) = \exp\left(-\tau^2 (2\pi\Delta f)^2 / 2\right). \quad (1)$$

where $\Delta f$ is the standard deviation the Gaussian envelop (and is related to the full width at half maximum (FWHM) by FWHM $\approx 2.355\Delta f$). $\Gamma(\tau)$ decreases exponentially with increasing $\tau$ until the fringes become vanishingly small. The coherence time $\tau_c$ of the laser is the optical delay needed to reduce the contrast to $1/e$. When $\tau >> \tau_c$ the MZI's average output power is independent of fluctuations in $\tau$. It is then highly stable against environmental changes, and consequently it produces little drift or noise.

The spectrum of the broadened laser with residual carrier of Fig. 2(a) is the sum of a broadband signal with normalized amplitude $f_g$ and linewidth $\Delta f_g$ and a narrow-linewidth signal with normalized amplitude $f_c$ and linewidth $\Delta f_c$, with $f_g + f_c = 1$. Applying the superposition principle to (1), the contrast of the output of the MZI probed with this laser is

$$\Gamma(\tau) = f_b e^{-\tau^2 (2\pi\Delta f_b)^2/2} + f_c e^{-\tau^2 (2\pi\Delta f_c)^2/2}. \quad (2)$$

Equation (2) states that as $\tau$ is increased, $\Gamma(\tau)$ decreases exponentially and asymptotically approaches a finite value equal to $f_c$. This is consistent with Fig. 2(b), which shows that for large $\tau$ the MZI signal does not go to 50% but continues to oscillate (up to a $\tau$ much larger than shown). In conclusion, the presence of residual carrier results in a wide range of path mismatch (corresponding to delays from ~50 fs to a much larger value) over which the MZI's output power is sensitive to length fluctuations. This is undesirable because for example in a LIDAR system it implies that a large number of scatterers contribute to noise and drift.

This analysis shows that the carrier extinction $f_c$ can be modeled simply as the contrast of an MZI at a large delay [13]. To derive an analytical expression of the contrast, start with the field of a laser modulated by a phase noise $\varphi_N(t)$:

$$E(t) = E_0 \exp\left(i\omega t + i\varphi_N(t)\right) \quad (3)$$

where $t$ is time and $\omega$ is the mean optical frequency. At the MZI output, the field interferes with a replica of itself delayed by $\tau$. The average normalized interferometric power is

$$\begin{aligned} \langle P(\tau)\rangle / P_0 &= \left\langle \left| E(t)/2 + E(t-\tau)/2 \right|^2 \right\rangle / E_0^2 \\ &= \left[1 + \cos(\omega\tau) E\left\{\cos\left(\varphi_N(t) - \varphi_N(t-\tau)\right)\right\}\right]/2 \\ &\quad - \sin(\omega\tau) E\left\{\sin\left(\varphi_N(t) - \varphi_N(t-\tau)\right)\right\}/2 \end{aligned} \quad (4)$$

where the expectation value (angled brackets) models the time averaging in the finite detection bandwidth. In general, the amplifier chain has approximately the same slew rate for positive and negative voltages. The expectation value of $\sin(\varphi_N(t) - \varphi_N(t-\tau))$ is then zero, and so is the last term in (4). In practice, the two slew rates differ a little, and this last term can have a nonzero value, but its contribution is generally negligible. Perfect symmetry in the amplifier slew rates is therefore assumed, because it provides a simpler model that fits the measured data well (see Sec. III). (4) then becomes

$$\langle P(\tau)\rangle = P_0\left(1 + \Gamma(\tau)\cos(\omega\tau)\right)/2. \quad (5)$$

with, since $\varphi_N(t) = \pi V_N(t)/V_\pi$,

$$\Gamma(\tau) = \left\langle \cos\left(\pi \frac{V_N(t) - V_N(t-\tau)}{V_\pi}\right)\right\rangle. \quad (6)$$

The value of $f_c$ can be determined by calculating $\Gamma(\tau)$ for $\tau$ larger than a few coherence times of the broadened laser [13]. In this limit, $V_N(t)$ and $V_N(t-\tau)$ are statistically independent. The PDF of $V_N(t) - V_N(t-\tau)$ is obtained by taking the auto-convolution of $P(V_N - \langle V_N\rangle)$ from which they were sampled, where $P(V_N)$ is the PDF of $V_N$ and $\langle V_N\rangle$ the mean of this PDF. Thus, the carrier suppression can be calculated by evaluating the expectation value of (6) using the convolution (*):

$$f_c = \left| \int_{-\infty}^{\infty} \cos\left(\frac{\pi V_N}{V_\pi}\right) \cdot P(V_N - \langle V_N\rangle) * P(V_N - \langle V_N\rangle) dV_N \right|. \quad (7)$$

(7) provides a simple expression for calculating the carrier suppression. The latter is only a function of the PDF of the applied phase shift, which can be derived from two parameters: $P(V_N)$ and $V_\pi$. No knowledge of the RF-frequency dependence is required to calculate the suppression, other than the average value of $V_\pi$ over the modulation bandwidth.

To evaluate it, two physical quantities are required: the value of $V_\pi$, which is known, and $P(V_N)$, which is generally not. This PDF can be either measured with an oscilloscope, as described in Sec. III, or it can be calculated using the method illustrated in Fig. 3. The PDF is calculated simply by converting the PDF of the Gaussian PDF of $v_n(t)$ through the nonlinear transfer function $h(v_n)$ of the amplifier. To this end, the required quantities are $p(v_n)$ and the inverse transfer function $h^{-1}(V_N)$. The PDF of $v_n$ is assumed to be a Gaussian, with a mean $v_\mu$ and a standard deviation $v_\sigma$. The reason why $v_\mu$ is not zero is that the first seven amplifiers add a small offset to the noise voltage. $P(V_N)$ is simply related to $p(v_n)$ by:

$$P(V_N) = p(h^{-1}(V_N))\left|\frac{d}{dV_N}h^{-1}\left(V_N\right)\right|. \quad (8)$$

As illustrated in the PDF shown in the top right corner of Fig. 3, the effect of saturation in the last amplifier is to distort the originally Gaussian input PDF. In principle the nonlinearity of the EOM's response also distorts the PDF, but it is far weaker than the nonlinearity of the amplifier response, and its impact on the model predictions is negligible.

Evaluating (8) requires the knowledge of the amplifier nonlinear transfer function. The latter has a sigmoidal shape that depends in its details on multiple physical parameters (e.g., a tanh function with polynomial arguments, as in [15]-[17]). In this work, for simplicity the transfer function was modeled with a tanh function but with fewer free parameters than in [15]-[17]:

$$\frac{V_N}{V_{sat}} = \begin{cases} x_a + (1+x_a)\tanh\left(\dfrac{Gv_n - V_{sat}x_a}{V_{sat}(1+x_a)}\right), & Gv_n < x_a V_{sat} \\ Gv_n / V_{sat}, & x_a V_{sat} \le Gv_n \le x_b V_{sat} \\ x_b + (1-x_b)\tanh\left(\dfrac{Gv_n - V_{sat}x_b}{V_{sat}(1-x_b)}\right), & Gv_n > x_b V_{sat} \end{cases} \quad (9)$$

where $v_n$ is the input voltage, and $G$ is the small-signal gain of the final amplifier.

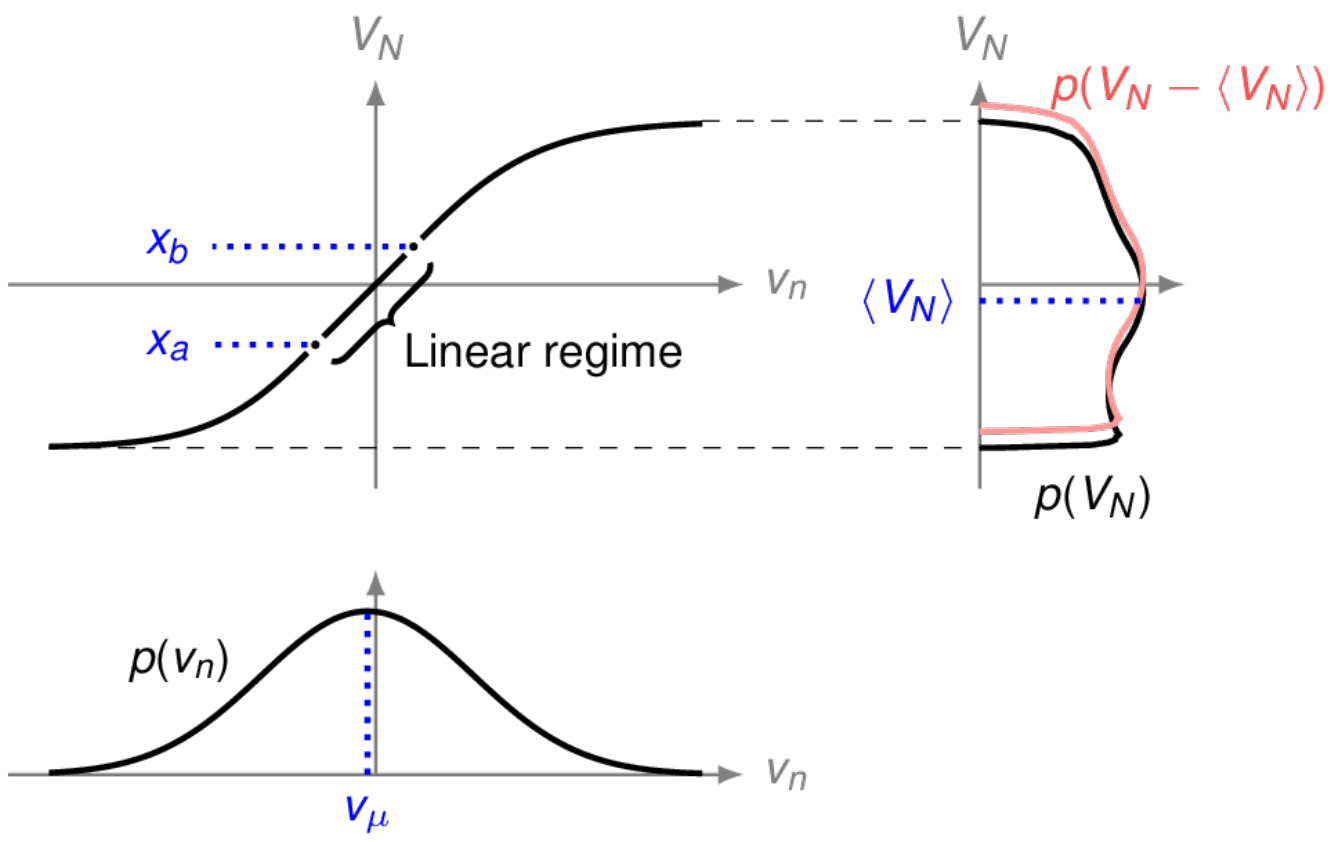

Figure 3. Diagram of the final amplifier's transfer function (upper left), the input PDF (lower left), and the output PDF (upper right).

In general, the transfer function of an RF amplifier is asymmetric because positive and negative voltages saturate slightly differently. This asymmetry is modeled in (9) by the parameters $x_a$ and $x_b$, which are the lower and upper bound of the amplifier's linear range, respectively. $x_a$ and $x_b$ take values between -1 and 1 and can be slightly different. This transfer function is plotted in Fig. 3 for the parameter values of the amplifier of Fig. 1, inferred from measurements reported in Sec. IIIA. The use of (9) is validated by the good quantitative agreement between the model predictions and several measurements described in that section.

To summarize, the carrier suppression is given by (7). For a given amplifier chain, (7) is evaluated by measuring the parameters $V_{sat}$, $G$, $x_a$, and $x_b$ of the last amplifier, as described in Sec. IIIA, which determines the transfer function $h(V_N)$ in (8); then measuring the $v_\sigma$ of the noise voltage applied to the last amplifier, which determine $v_n(t)$ and therefore its PDF $p(v_n)$. $h^{-1}(V_N)$ and $p(v_n)$ are then used in (8) to compute $P(V_N)$, which is inserted in (7) to obtain the carrier suppression.

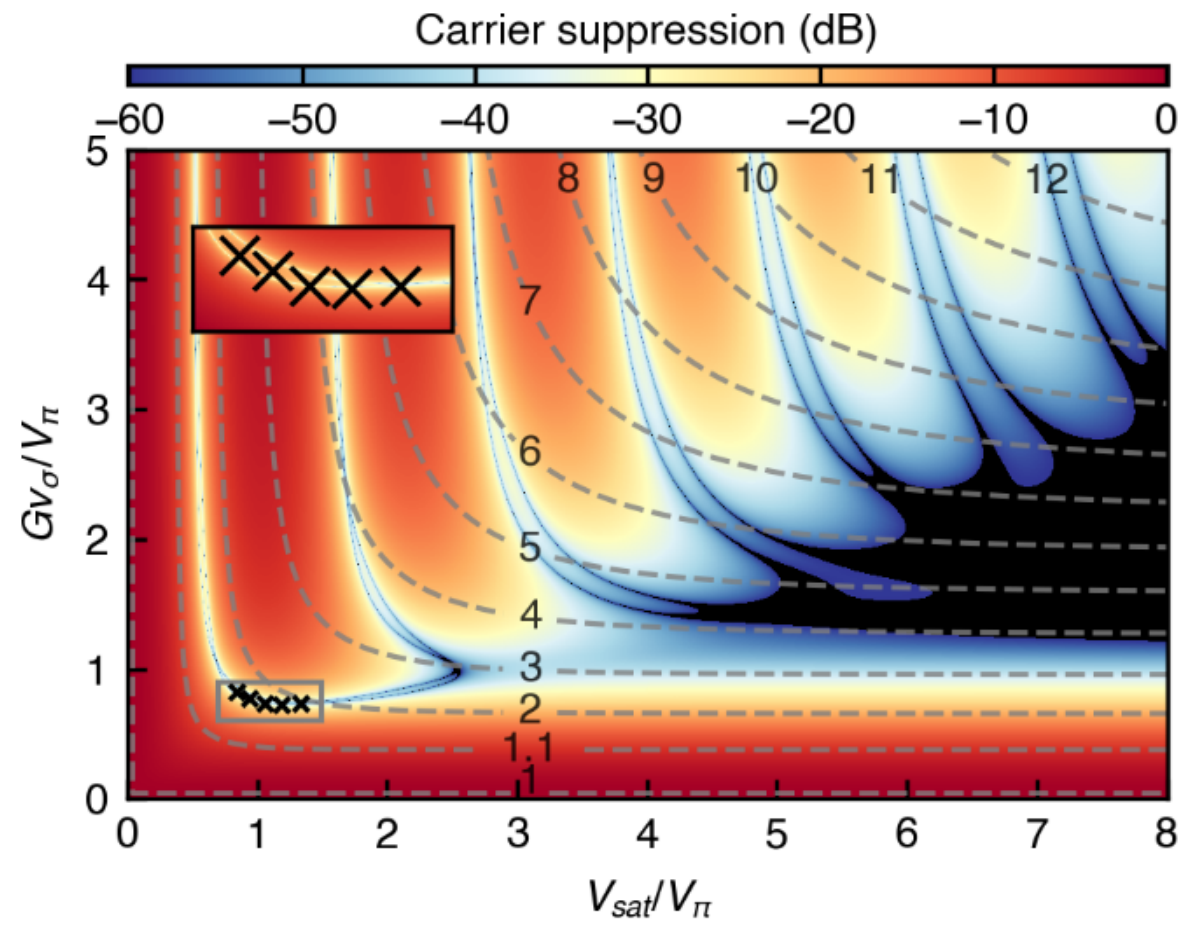

Figure 4. Heat map of the carrier suppression $f_c$ as a function of $Gv_\sigma/V_\pi$ and $V_{sat}/V_\pi$, assuming an amplifier transfer function (9) with $G$ = 18 dB, $x_a$ = -0.356, $x_b$ = 0.148, input PDF $p(v_n)$ with $v_\mu$ = -40.9 mV.

Figure 4 plots the dependence of the carrier suppression on $Gv_\sigma$ and $V_{sat}$ calculated by solving (7) for the values of $G$, $x_a$, and $x_b$ measured for the last amplifier of Fig. 1. The vertical axis is $Gv_\sigma/V_\pi$, and the horizontal axis $V_{sat}/V_\pi$. Poor carrier suppression (-20 dB and above) is shown in red, excellent suppression (-60 dB and below) in black. At the very bottom ($Gv_\sigma = 0$, no modulation) there is no carrier suppression (deep red). Maximum suppression occurs in thin twin bands. The first set of bands, located between $V_{sat}/V_\pi$ of ~0.56 and ~2.56, has the shape of a distorted U, distorted in the sense that its bottom right is pulled toward larger $V_{sat}$ values. The second, third, and higher sets are also U-shaped bands that are even more distorted, to the point where they merge into a wide region of excellent carrier suppression (the large black region for $V_{sat}/V_\pi$ greater than ~6 in Fig. 4). The level of distortion and the shape of the bands depend critically on how smoothly the amplifier saturates, namely on $x_a$ and $x_b$. If $-x_a = x_b = 1$ (i.e., the amplifier is linear from $-V_{sat}$ to $+V_{sat}$), the distortion is weak, and each set of bands has a distinct U shape. The reason why the bands come in pairs is the asymmetry in $P(V_N)$, which allows two nearby values of $Gv_\sigma$ to result in PDFs that perfectly suppress the carrier. When $x_a = -x_b$ and $v_\mu = 0$, the pairs merge together.

When $Gv_\sigma >> V_{sat}$ (top left of Fig. 4), the final amplifier is almost always saturated. All the power is then in the two side lobes, which can be approximated as delta functions at $\pm V_{sat}$, i.e., $P(V_N) = (\delta(V_N - V_{sat}) + \delta(V_N + V_{sat}))/2$. Substituting this equation in (7) yields $f_c = (1 + \cos(\pi V_{sat}/V_\pi))/2$, which agrees with [18], and describes the periodic structure along the top of Fig. 4: the carrier suppression is infinite when $V_{sat}/V_\pi = (2n + 1)/2$, where $n$ is an integer.

When $Gv_\sigma << V_{sat}$ (bottom right in Fig. 4), the PDF is approximately Gaussian. Inserting a Gaussian into (7) yields $f_c = \exp(-(\pi Gv_\sigma/V_\pi)^2)$, which agrees with [11]-[14]. This expression describes the roll-off into a wide region of excellent carrier suppression at the far right of Fig. 4.

The dashed curves in Fig. 4 are contours of equal broadening, calculated with the model presented in Appendix A. The number on each contour is to the ratio of the bandwidth (standard deviation) of the broadened light to the electrical bandwidth (standard deviation) of the noise voltage at the EOM. These curves show that except for a very low $Gv_\sigma$ or $V_{sat}$ (which hardly ever occurs in practice), this ratio is larger than 1: *the broadened light is broader than the noise signal.* For example, along the curve labeled "4," if the amplifier chain has a bandwidth of 10 GHz the broadened laser has a bandwidth of 40 GHz. The laser is broadened to 40 GHz with only 10-GHz electronics, which represents a significant saving in complexity and cost. As $V_{sat}$ and $Gv_\sigma$ are increased, this ratio increases.

In some applications, a finite power budget or thermal considerations limit the practical maximum value of $Gv_\sigma$. The practical operating points that require the least power while perfectly suppressing the carrier are along the bottom of the first band in Fig. 4. This region has the lowest $Gv_\sigma$, but at the cost of tight tolerances in both $Gv_\sigma$ and $V_{sat}$. For example, Fig. 4 shows that for a $V_\pi$ of 4.7 V and a $V_{sat}$ of 6.3 V ($V_{sat}/V_\pi = 1.34$), -30 dB of suppression can be achieved with $Gv_\sigma/V_\pi = 0.73 \pm 0.03$.

Operating in a higher order band will relax these tolerances, because these bands are much wider. A region of particular interest is the broad dark-blue band that starts at $V_{sat}/V_\pi$ of ~5. This region can be reached with a $V_\pi$ of 3.5 V and a $V_{sat}$ of 21 V. A suppression of -30 dB can now be achieved with $Gv_\sigma/V_\pi \approx 3.17 \pm 2.33$: the tolerance is ~70 times greater. The tolerance on $V_{sat}/V_\pi$ is also much larger. An amplifier with such a large $V_{sat}$ will have a lower bandwidth, but in this region the ratio of optical to electrical bandwidth is higher (~4.4) than at the bottom of the first band (~2) (see Fig. 4), which is a significant compensation. The combination of a low-$V_\pi$ EOM such as available from EOSPACE ($V_\pi = 3.5$ V), and an amplifier such as the ZHL-5W-202-S+ from Minicircuits, with a 3-dB compression point at 38 dBm (resulting in a $V_{sat}$ of ~21 V) and a bandwidth of 2.2 GHz, would be able to reach this region of high tolerance in $V_{sat}$, $V_\pi$, and $Gv_\sigma$, and provide the required optical bandwidth (a FWHM of ~22.7 GHz). The downside is increased cost.

## III. Experimental Verification

### A. PDF and Transfer Function Measurements

In a first set of measurements, the PDF of the noise voltage $V_N(t)$ at the output of the amplifier chain in Fig. 1 was measured. The objectives were (1) to verify that the dependence of the PDF's shape on $Gv_\sigma$ matched the model, and (2) to infer from this dependence the last amplifier's transfer function (i.e., the values of $V_{sat}$, $G$, $x_a$, and $x_b$). These values are needed to generate, for this amplifier, the heat map of Fig. 4, which predicts the values of $Gv_\sigma$ and $V_{sat}$ for optimum carrier suppression. The PDF and corresponding periodogram were measured with an Agilent 86100A 20-GHz oscilloscope (fast enough to resolve the ~10 GHz modulation noise) for six values of $v_\sigma$ by varying the attenuation $\alpha_1$ while keeping $\alpha_2$ constant at 0 dB (see Fig. 1). The measured PDF distributions are shown in Fig. 5 (solid blue curves). The periodograms were approximately flat (±0.25 dB) from 150 MHz to 10 GHz.

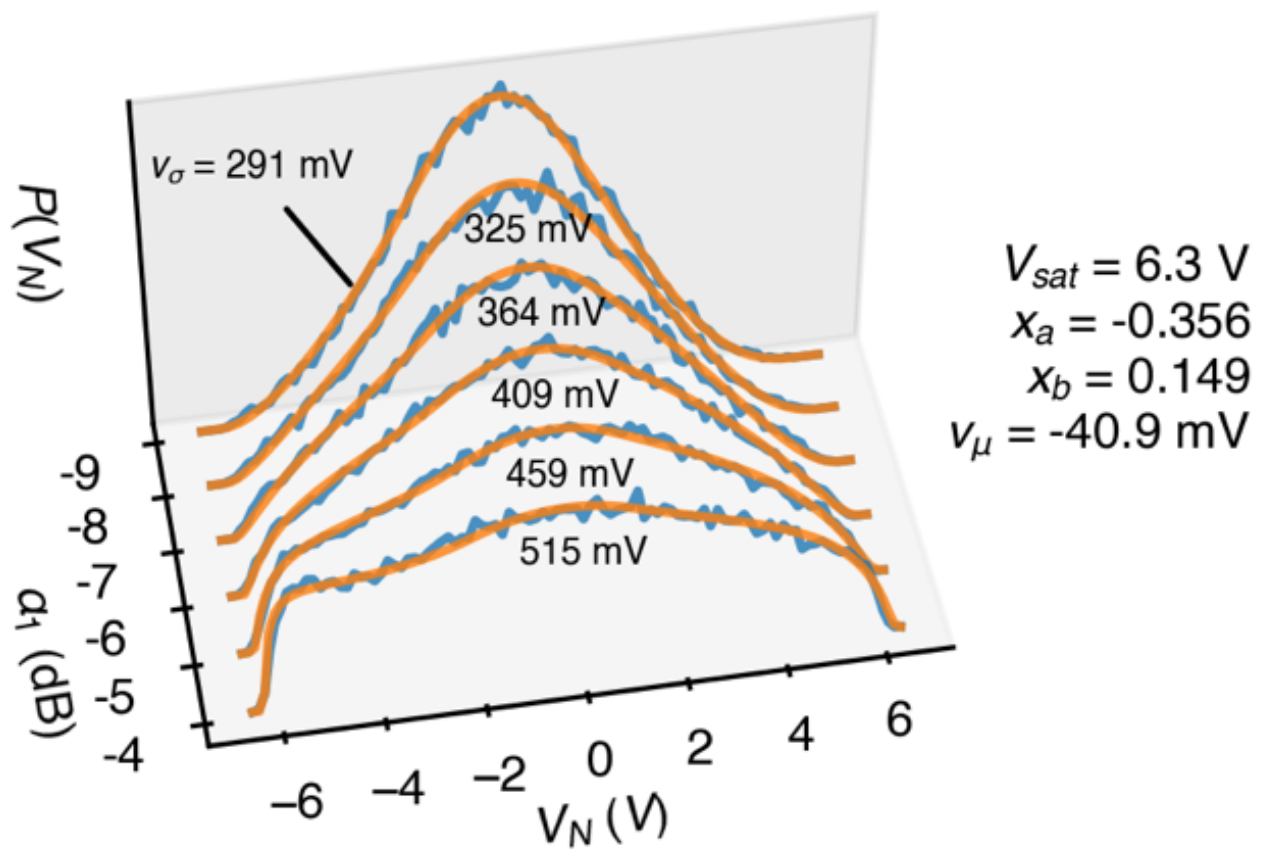

Figure 5: Output PDFs of the amplifier chain of Fig. 1 measured at various attenuations of the input noise $\alpha_1$ (solid curves) and best fits (dashed curves).

As $\alpha_1$ was tuned from -9 dB to -4 dB, the PDF widened (i.e., the voltage increased). Because the transfer function saturates at $V_N = \pm V_{sat}$, the output PDF is confined between $\pm V_{sat}$, causing the distribution to also become distorted: it flattened out and developed lobes near $\pm V_{sat}$ (most apparent on the left side of the curve for $\alpha_1 = -4$ dB). The measured PDFs were fitted to (8) using $G = 18$ dB for the gain (value provided by the manufacturer) and performing a least-square regression on the amplifier parameters $V_{sat}$, $x_a$, and $x_b$, and the mean of the input Gaussian noise $v_\mu$. The fits (orange curves in Fig. 5) match the measured PDFs very well. The parameter values that gave these best fits are listed in Fig. 5. The three fitted values for the amplifier ($V_{sat}$, $x_a$, and $x_b$), which entirely define the transfer function, yield a 1-dB compression power for this amplifier for a 50-Ω load of 25.1 dBm. This is very close to the typical value of 25 dBm provided by the manufacturer.

### B. Carrier Suppression Measurements

When a FOG is interrogated by a narrow-linewidth laser, the output-signal noise is limited by backscattering in the FOG's fiber coil, which form many spurious interferometers [6]. When the laser is broadened, the residual carrier power decreases, which reduces the noise in proportion to this residual power [10]. The carrier suppression can therefore be inferred by comparing the measured noise of a FOG with and without laser broadening.

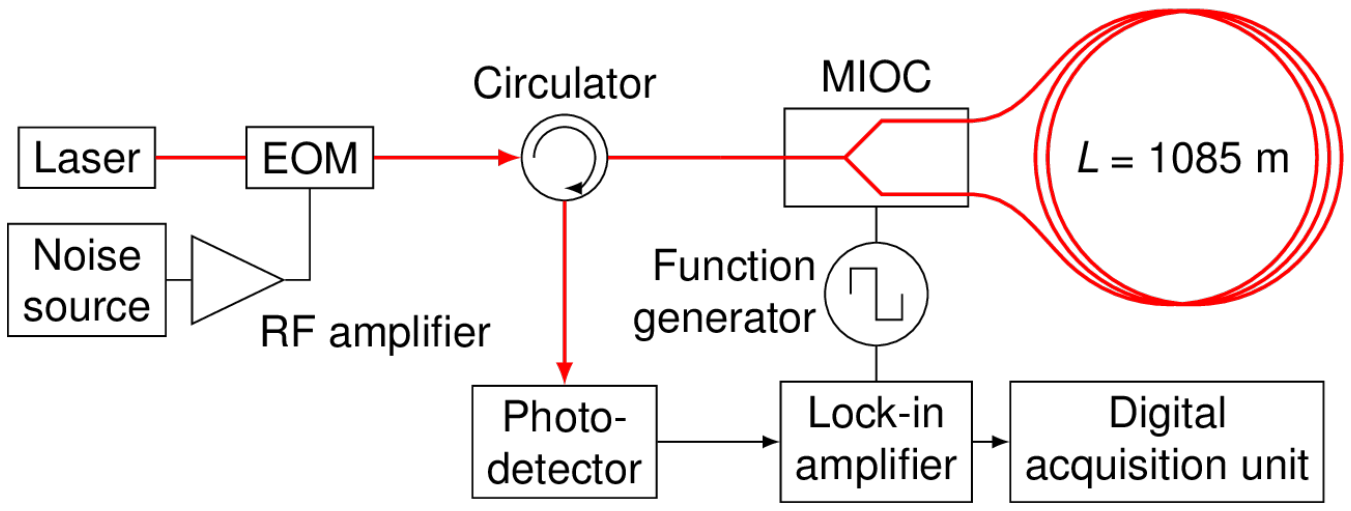

Figure 6. Schematic of the broadened-laser-driven FOG used to measure coherence suppression.

The FOG configuration, described in detail in [10], is shown in Fig. 6. In brief, light from a laser (linewidth of ~90 Hz) was passed through a phase modulator driven by the noise source of Fig. 1. The broadened light was sent through a circulator and into a multi-function integrated-optic circuit (MIOC), which polarizes the light, splits it evenly into the two inputs of the fiber coil, and applies to them a square-wave phase modulation to bias the interferometer for maximum signal-to-noise ratio. The two signals returning from the coil are recombined in the MIOC and directed by the circulator to a photodetector. The photodetector signal is demodulated with a lock-in amplifier, whose output (proportional to the rotation rate of the FOG) is measured with a digital acquisition system.

The carrier suppression of the broadened laser was measured by recording the output signal of this FOG for a few minutes, calculating the Allan variance of this temporal trace, and inferring the noise from the Allan variance. This measurement was carried out first with the unbroadened laser, then with the same laser broadened at various $Gv_\sigma$ values in order to vary $f_c$ (while keeping $\alpha_2 = 0$ dB (see Fig. 1) so that $V_{sat}$ was constant and equal to 6.3 V). The $f_c$ values inferred from this measurement are plotted in Fig. 7 as orange dots. As $Gv_\sigma$ was increased, the carrier suppression first improved rapidly to a minimum of -37 dB at ~3.4 V, then it degraded. This behavior is identical to the findings of [9], [10].

Figure 7 also plots in blue the carrier suppression either calculated using (7) (dashed curve) or from the PDFs measured at various $Gv_\sigma$ values with an oscilloscope (as done in Fig. 5(a)) (blue circles). $V_\pi$ is RF-frequency dependent. The EOM manufacturer lists a nominal $V_\pi$ of 4–5 V at 50 kHz, and 6–7 V at 10 GHz. The value of $V_\pi = 4.7$ V used in these simulations, consistent with these values, gave a good agreement between the orange dots and blue circles. The modeled and measured dependencies of the carrier suppression on $Gv_\sigma$ also agree well. This confirms that the narrow range of high carrier suppression around $Gv_\sigma = 3.43$ V is caused by saturation of the amplifier: without it, as predicted in [13], [14], the carrier suppression would keep improving as $Gv_\sigma$ is increased.

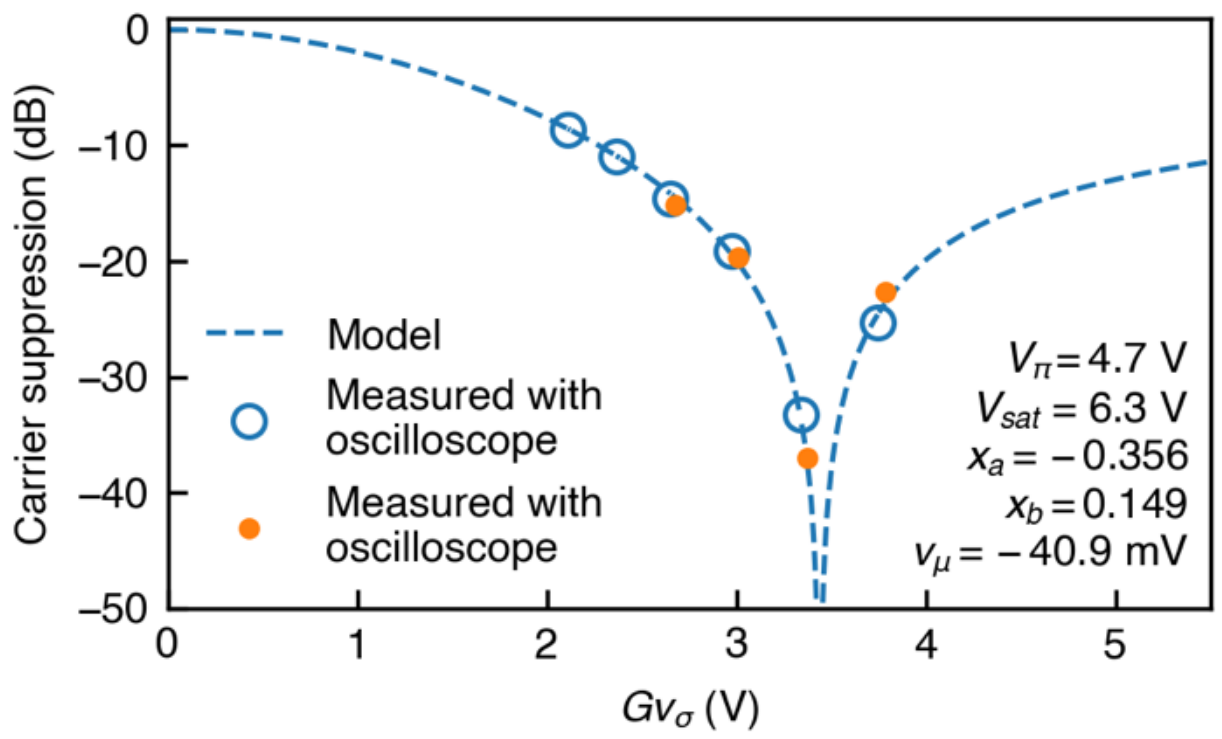

Figure 7. Model for carrier suppression plotted against measured carrier suppression in a FOG and inferred carrier suppression from oscilloscope trace.

The minimum in Fig. 7 corresponds to a point near the bottom of the first band in Fig. 4. To confirm the shape of the first band, the noise was measured again, but this time for different attenuations $\alpha_1$ and $\alpha_2$ (see Fig. 1) to vary both $Gv_\sigma$ and $V_{sat}$. The combinations of $Gv_\sigma$ and $V_{sat}$ that minimized the noise, and therefore $f_c$, are plotted as crosses in Fig. 4. These points agree very well with the theoretical band. Operating at these points gave a drift of 5.5 mdeg/h, a current record for a laser-driven FOG thanks to the predictions of this model.

Operating on the right side of the first band in Fig. 4 would be advantageous because the carrier suppression is less sensitive to variations in $V_{sat}$ and $Gv_\sigma$. To reach that region, the electronics and EOM need to satisfy $V_{sat}/V_\pi = 2.5$ (see Fig. 4). To achieve a ratio of 2.5 or greater would require an EOM fabricated in a longer electro-optic crystal to reduce $V_\pi$, and/or an amplifier that can supply larger voltages. Both enhancements come at the cost of a reduced noise bandwidth, and therefore a reduced optical bandwidth, and an increased temperature sensitivity of the EOM.

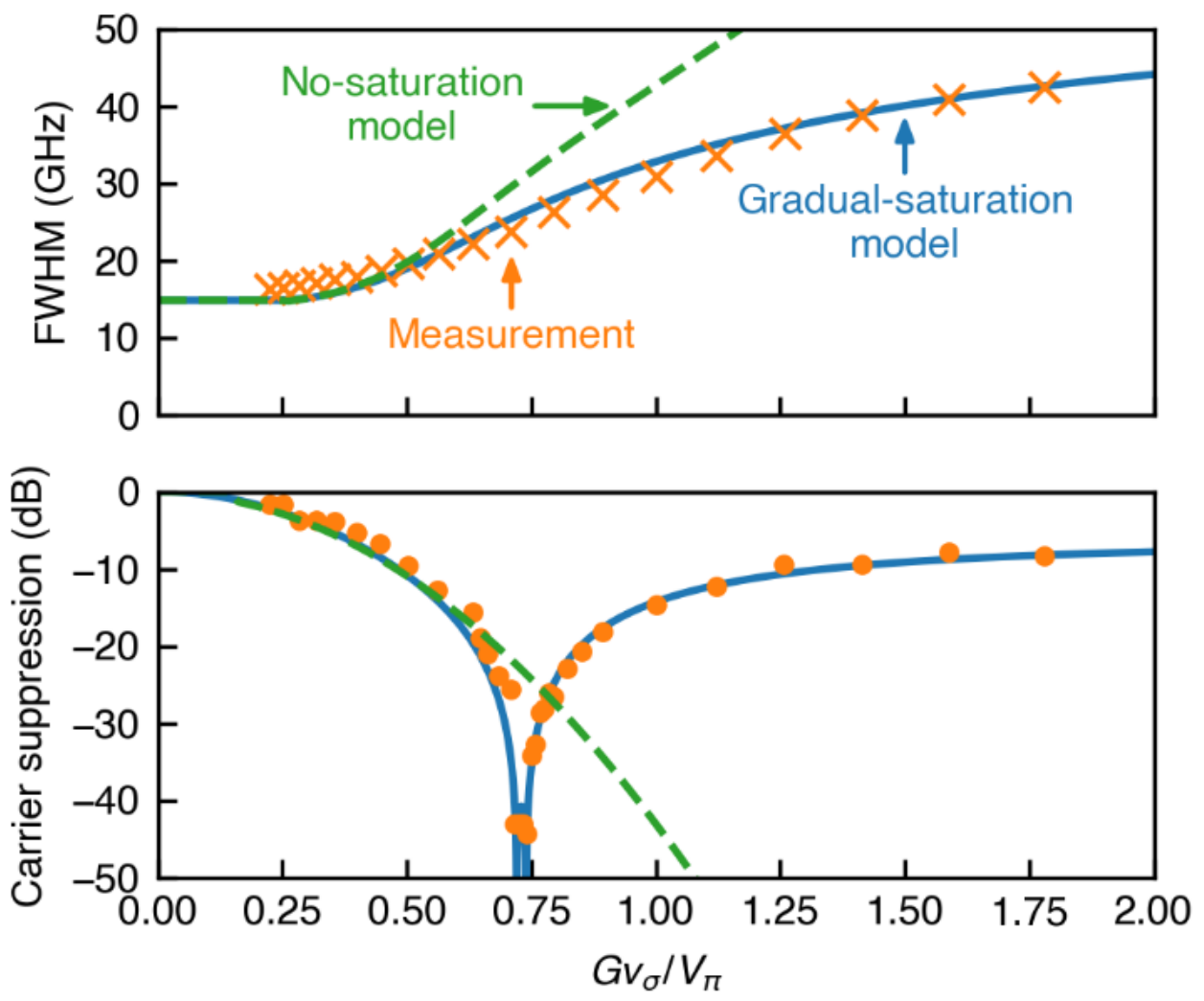

Figure 8. Measured FWHM and carrier suppression of the broadened laser of [10], and mathematical predictions of this work. Model parameters are based on device parameters as listed by the EOM and amplifier manufacturers.

### C. Broadened Linewidth Measurements

The dependence of the linewidth of a broadened laser on $Gv_\sigma$ was measured with the same FOG interrogated with a 2-kHz-linewidth RIO laser broadened with a slightly different noise source [10] (the front end in Fig. 1 was replaced by a commercial noise source, but the final amplifier was nominally the same). The carrier suppression was measured with two different methods. The first one was identical to the Allan-variance method described in relation to Fig. 7, except that the carrier suppression was inferred from the measured drift instead of the noise, as reported in [10]. The drift was defined as the minimum of the measured Allan-variance curve. In the second method, the carrier suppression was obtained by measuring the broadened spectrum with an OSA. The spectrum was then fitted to a Gaussian to recover the residual carrier (the height of the bump at the top of the spectrum, see top right of Fig. 1). This fitting process also provided the linewidth of the broadened laser.

The results of are plotted in Fig. 8. The linewidth increases monotonically as $Gv_\sigma$ is increased (orange crosses in Fig. 8(a)). The linewidth curves were calculated with the model of Appendix A. The curve with saturation was calculated for the $V_{sat} = 4.7$ V of the last amplifier. The curve without saturation was calculated with the same model assuming an infinite $V_{sat}$. The model accurately matches the measured dependence of the broadened-laser linewidth on $Gv_\sigma$. Comparison between the dashed and solid curves shows that saturation slows down the increase in linewidth (Fig. 8(a)).

The carrier suppression inferred from the FOG's measured drift (orange dots in Fig. 8(b)) exhibits a similar dependence as in Fig. 7, with a best suppression of -44 dB. The theoretical curves were calculated for the inferred transfer function of the last amplifier (solid curves) and for a hypothetical amplifier with no saturation (dashed curves). Here too there is excellent agreement between the measured data and the predicted curve in the presence of saturation. Comparison between the dashed and solid curves in Fig. 8(b) shows that saturation reduces the $Gv_\sigma$ required to achieve a high carrier suppression.

## IV. Conclusion

A simple analytical expression was derived for the carrier suppression in a laser broadened by an arbitrary random phase modulation waveform, with a focus on Gaussian white noise distorted by practical, saturated amplifiers. It also provides an approximate expression for the lineshape of the broadened optical spectrum. Conditions are derived for the voltage $Gv_\sigma$ applied to the EOM, the saturation voltage $V_{sat}$ of the amplifier chain, and the $V_\pi$ of the EOM required for maximum carrier suppression. The model can be implemented regardless of whether the final amplifier is linear over all or only part of its output range. The conditions required to achieve maximum carrier suppression predicted by this model are validated by measuring the suppression of noise and drift in a FOG driven by such a broadened laser. A maximum carrier suppression of -44 dB was measured with a FWHM of 27 GHz.

The model helps identifying regions of optimum carrier suppression with large tolerances and predicting the linewidth at those settings. For example, for a $V_\pi$ of 4.7 V and a $V_{sat}$ of 6.3 V, -30 dB of carrier suppression is predicted by the model with $Gv_\sigma/V_\pi = 0.73 \pm 0.03$. For these settings, the optical bandwidth is more than twice the electrical noise bandwidth. Even greater tolerances and linewidths are predicted with amplifiers that supply higher voltages. Surprisingly, amplifier saturation allows perfect carrier suppression at lower voltages than possible with amplifiers that do not saturate, which reduces the power requirement at the expense of tighter voltage tolerances.

## Acknowledgment

The authors thank Joseph Kahn for the use of his wide-band oscilloscope, attenuators, and amplifiers.

## Appendix A: Derivation of the Broadened Linewidth

Prior investigations [11]-[13] reported a numerical method to calculate the lineshape of a broadened laser in the transition region between weak modulation (the broadened laser then has a top-hat lineshape) and strong modulation (Gaussian lineshape) with no saturation. In this appendix, an analytical approximation of the lineshape is derived for any modulation strength while accounting for saturation. The broadband noise voltage $V_N(t)$ of bandwidth $\omega_{max}$ is approximated as a Fourier sum of a large number $N$ of equally spaced discrete frequencies. Each component has a frequency $\omega_i$ (spaced by $\omega_{max}/N$) and a random phase $\varphi_i$. Absent any saturation, the standard deviation of $V_N(t)$ is $Gv_\sigma$, and the rms voltage of each of the discrete frequency components is $Gv_\sigma/\sqrt{N}$. Because the peak value of a sine is $\sqrt{2}$ times its rms value, $V_N(t)$ can be expressed as

$$V_N(t) = \sum_{i=1}^{N} Gv_\sigma \sqrt{2/N} \sin\left(\omega_i t + \varphi_i\right). \qquad (10)$$

An EOM driven by broadband noise is modeled as a series of $N$ EOMs, each driven at a single frequency $\omega_i$ and producing sidebands at $\pm\omega_i$. Each EOM provides the same phase-modulation amplitude $\Phi_p = (2/N)^{1/2} \pi Gv_\sigma/V_\pi$. The optical spectrum $S(\omega)$ of the output of an EOM after phase modulation from frequency $\omega_i$ is given by the Jacobi-Anger expansion:

$$S(\omega) = P_0 \sum_{m=-\infty}^{m=\infty} J_m^2\left(\Phi_p\right) \delta\left(\omega - m\omega_i - \omega_0\right). \qquad (11)$$

where $P_0$ is the output power and $\omega_0$ the carrier frequency. For weak modulation, $Gv_\sigma$ is small, and so is $\Phi_p$. Only the terms in $|m| \le 1$ are significant. $J_0^2(\Phi_p) \approx 1$, and $J_{\pm 1}^2(\Phi_p) \approx \Phi_r^2/(2N)$, where $\Phi_r = \pi Gv_\sigma/V_\pi$. The output optical spectrum of the first EOM contains the $\omega_0$ component with power $\sim P_0$, and two sidebands at $\omega_0 \pm \omega_1$ each with a power $P_0\Phi_\sigma^2/2N$.

The second EOM adds sidebands at $\pm\omega_2$ for each discrete frequency output by the first EOM. Thus, its output contains the $\omega_0$ component with power $\sim P_0$, the unmodified sidebands from the first EOM at $\omega_0 \pm \omega_1$ with power $P_0\Phi_\sigma^2/2N$, the sidebands of the carrier created by the second EOM at $\omega_0 \pm \omega_2$ with power $P_0\Phi_\sigma^2/2N$, and sidebands of sidebands at $\omega_0 \pm \omega_1 \pm \omega_2$ with power $P_0\Phi_\sigma^4/4N^2$, which are negligible.

By induction, the output of the last EOM can be approximated as a central carrier with power $f_cP_0$ (where $f_c$ is determined using the methods of Sec. III), surrounded by a frequency comb of sidebands each with power $P_0\Phi_\sigma^2/2N$. Since $N$ is large, this comb can be written as a continuous top-hat distribution with total power $P_0\Phi_\sigma^2$ and half-width $\omega_{max}$. A flat RF spectrum is assumed for simplicity. However, this approach can also model an arbitrary RF noise spectrum and EOM frequency response. In this more general case, the sidebands will have different powers given by $P_0\Phi_i^2/2N$, where $\Phi_i = \pi G v_{\sigma,i}/V_\pi(\omega_i)$ and $Gv_{\sigma,i}$ is determined by the RF power in the band $(\omega_i, \omega_i + \omega_{max}/N)$.

As $Gv_\sigma$ is increased (still in the limit of no saturation), the weak-modulation approximation no longer holds, and the spectral shape evolves. Specifically, each first-order sideband becomes attenuated by $J_0^2(\Phi_p)$ (which is no longer ~1) at each EOM and the power in the sidebands of sidebands is no longer negligible. The output spectrum is then obtained by convolving the spectrum of these sidebands with each other (proposed for weak modulation in [13] and validated for strong modulation in [12]). The power remaining in the carrier is then accurately modeled by (7) to describe $f_c$. After passing through $N$ EOMs, the power in the top-hat distribution (which consists entirely of sidebands of the central carrier) is reduced by a factor of $J_0^2(\Phi_p)$ a total of $N$ times. The power remaining in the top-hat distribution is then

$$\lim_{N\to\infty}\left(J_0^2\left(\Phi_p\right)^N\right) = \lim_{N\to\infty}\left(J_0^2\left(\Phi_\sigma\sqrt{2/N}\right)^N\right) = e^{-\Phi_\sigma^2}. \quad (12)$$

The rest of the power is modeled as repeated convolutions of sidebands of sidebands. The central limit theorem states that repeated convolutions of PDFs yield a Gaussian spectrum whose variance is the sum of the variance of each constituent PDF. Hence, the fraction of power in the carrier is $f_c$, in the top hat is $f_t = \Phi_\sigma^2\exp(-\Phi_\sigma^2)$, and in the Gaussian $f_g = 1 - f_t - f_c$.

To calculate the linewidth of the Gaussian distribution, the variances of the spectra generated by phase modulation of a single input frequency by each discrete frequency component are simply added. The variance $\sigma_i^2$ of the Jacobi-Anger expansion given in (11) for a single frequency $\omega_i$ is

$$\sigma_i^2 = \sum_{m=-\infty}^{m=\infty} J_m^2\left(\Phi_p\right)\left(m\omega_i\right)^2 = \frac{\Phi_p^2}{2}\omega_i^2 = \frac{\Phi_\sigma^2}{N}\omega_i^2. \quad (13)$$

Summing these variances gives:

$$\sigma^2 = \sum_{n=1}^{N}\frac{\left(n\omega_{max}/N\right)^2}{N}\Phi_\sigma^{\ 2} = \frac{\omega_{max}^2}{3}\Phi_\sigma^{\ 2} + O\left(N^{-1}\right). \quad (14)$$

When saturation is present, the derivation is the same except for how $Gv_\sigma$ is determined. Numerical simulations show that saturation reduces all components of the Fourier series equally by a factor $Gv_\sigma/V_\sigma$, where $V_\sigma$ is the true standard deviation of the compressed PDF.

The total broadened optical spectrum is the superposition of the residual carrier with amplitude $f_c$, a top-hat lineshape of half-width $\omega_{max}$ and amplitude $f_t$, and a Gaussian lineshape with a standard deviation $\omega_{max}\pi V_\sigma/V_\pi/\sqrt{3}$ and amplitude $1 - f_c - f_t$, where $f_t$ is calculated using $V_\sigma$ instead of $Gv_\sigma$. A graphical representation of this superposition is shown in Fig. 6.2 in [11]. 98% of the power is contained within a bandwidth of $\pm\omega_{max}\max(0.98, 2\Phi_\sigma/\sqrt{3})$, which agrees with the well-known Carson's bandwidth approximation [20].